\documentclass[10pt,conference]{IEEEtran}
\usepackage{bm}
\usepackage{mathrsfs}
\usepackage{textcomp}
\usepackage{epsfig}
\usepackage{indentfirst}
\usepackage{amsmath}
\usepackage{amssymb}
\usepackage{array}
\usepackage{multirow}
\usepackage{graphicx}
\usepackage{subfig}

\usepackage{cite}
\begin{document}
\title{Unequal Error Protected JPEG 2000 Broadcast Scheme with Progressive Fountain Codes}
\author{{Zhao Chen$^1$, Mai Xu$^1$, Luiguo Yin$^2$ and Jianhua Lu$^1$}\\
$^1$Department of Electronic Engineering, Tsinghua University\\
$^2$School of Aerospace, Tsinghua University\\
Email: zhao-chen10@mails.tsinghua.edu.cn; \{xumai, yinlg, lhh-dee\}@tsinghua.edu.cn}%@mail.tsinghua.edu.cn

\maketitle

\begin{abstract}
This paper proposes a novel scheme, based on progressive fountain codes, for broadcasting JPEG 2000 multimedia. In such a broadcast scheme, progressive resolution levels of images/video have been unequally protected when transmitted using the proposed progressive fountain codes.
With progressive fountain codes applied in the broadcast scheme, the resolutions of images (JPEG 2000) or videos (MJPEG 2000) received by different users can be automatically adaptive to their channel qualities, i.e. the users with good channel qualities are possible to receive the high resolution images/vedio while the users with bad channel qualities may receive low resolution images/vedio. Finally, the performance of the proposed scheme is evaluated with the MJPEG 2000 broadcast prototype.
\end{abstract}

\section{Introduction}\label{Intro}
Broadcast \cite{Wieselthier2000} offers the promise of overcoming the bandwidth limitation in wireless communications by using one channel to transmit data to all users. Therefore, in many communication networks broadcast is often desired and required. For example, in 3G UMTS cellular networks, Multimedia Broadcast Multicast Service (MBMS) \cite{3GPP_MBMS} has been proposed as a standard of 3GPP for providing multimedia service to users via broadcast.

In a broadcast network, multimedia plays a key role as it is the content to be transmitted, and it mainly includes the forms of text, audio, images and video. As the state-of-art image compression standard, Joint Photographic Experts Group 2000 (JPEG 2000) \cite{Rabbani2002} is pervasive in broadcast network since it is capable of providing efficient image or even video (Motion JPEG 2000, MJPEG 2000) information content to the users. JPEG 2000 has already been issued  recently  as the standard by ISO/IEC 15444 \cite{J2000} for supporting lossy and lossless compression of images. The success of JPEG during the past decade also implies a wide application of JPEG 2000 in communication systems in the future. One of the most attractions of JPEG 2000 is that it is able to produce progressive recovery of an image by fidelity or resolution.

Besides, the wireless transmission techniques are also important for a broadcast network since they offer the carrier for multimedia content. Fountain codes (also known as rateless erasure codes) is one of such wireless transmission technique in which the original source symbols can be accurately recovered from any subset of the encoding symbols with the size equal to or only slightly larger than the number of source symbols.
Luby Transform (LT) codes\cite{luby2002lt} or Raptor codes\cite{shokrollahi2006raptor}, as two state-of-art techniques of fountain codes, have been proved to be an efficient forward error correction (FEC) solution for erasure channels\footnote{FEC gives the receiver an ability to correct errors without data retransmission. Over an erasure channel, the receiver either receives the packet or drop it when error is detected.}. These codes are universal for different scenarios on packet transmission level regardless of channel packet loss patterns. So, fountain codes are becoming increasingly popular in broadcast network. For example, Raptor codes, with nearly linear encoding/decoding complexity, have been accepted for the application layer FEC scheme in current communication standards, such as 3GPP MBMS\cite{3GPP_MBMS} and DVB-H\cite{DVB_H}.

In JPEG 2000 broadcast network\footnote{ Note that video can also be transmitted in this broadcast network using MJPEG 2000, each frame of which can be seen as JPEG 2000 images.}, scalable image/video transmissions can be achieved once considering the progressive levels of images in JPEG 2000 format or the progressive levels of frame images in MJPEG 2000 format.
However, when applying fountain codes to scalable image/video transmissions in such a broadcast network, unequal error protection (UEP) strategies have to be considered since different levels of images/video have different priorities. Recently, there are several UEP methods proposed for scalable image/video streaming.
In \cite{cataldi2010sliding}, \cite{sejdinovic2009expanding} and \cite{hellge2008multidimensional}, data of each priority level are encoded and decoded by different sets of fountain codes separately, leading to rather high complexity and large overhead. Another method in \cite{rahnavard2007rateless} changes the degree distribution in order to encode the layered stream data together with UEP ability, but it decreases the decoding efficiency resulting in a larger overhead. In \cite{chang2008unequal}, hierarchical coding graphs are proposed to obtain different decoding paths for layered encoded symbol, but the overhead is still high due to a fixed structure of the coding graph.

\begin{figure*}
\centering
\includegraphics[height=7cm]{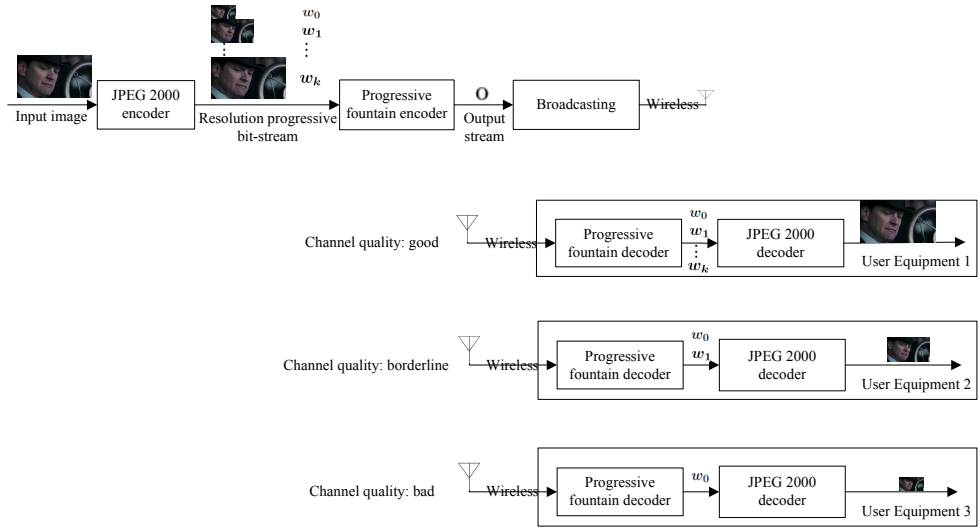}
\caption{Framework of the proposed unequal error protected JPEG 2000 broadcast scheme with progressive fountain codes. Note that the notations in figure will be discussed in the following sections.}
\label{Framework}
\vspace{-1.5em}
\end{figure*}

In this paper, for scalable image/video stream transmission, a progressive fountain codes is proposed to apply UEP to the broadcast of each progressive level information of JPEG 2000 images, with high efficiency and low complexity. Beyond the proposed progressive fountain codes, unequal error protected JPEG 2000 broadcast scheme is designed with the purpose that different users can receive broadcasting images/videos with different resolutions according to their channel qualities. The framework of such JPEG 2000 broadcast scheme can be seen in Figure \ref{Framework}.

The rest of this paper is organized as follows. In Section \ref{JPEG2000} we show some basic concepts of progressive resolution recovery of JPEG 2000. Section \ref{ProFC} contains our proposed progressive fountain codes for unequal error protected image/video transmission. The prototype of the proposed broadcast scheme is introduced in Section \ref{System}. Section \ref{Result} shows the experimental results and compares our scheme with EEP schemes. Finally, we conclude the paper in Section \ref{conclusion}.

\section{Progressive Resolution Recovery of JPEG 2000}\label{JPEG2000}
One of most attractive characteristics of JPEG 2000 is that the progressive resolution recovery of compressed images, as a multi-resolution image representation, is inherent to Discrete Wavelet Transform (DWT) that is a key algorithm of JPEG 2000 standard. In this section, we shall concentrate on the decomposition technique of DWT, as the foundation of progressive resolution recovery in JPEG 2000.

Assume $\mathbf{I}$ to be the input image to the compression algorithm of JPEG 2000. Let us first consider the one-dimensional DWT (1-D DWT).  In 1-D DWT, the samples  of one-dimensional signal $\mathbf{x}$ are passed through a low-pass filter $h_0$ and downsampled by a factor of two:
\begin{equation}
y_{low}[m] = \sum_{l=-\infty}^{+\infty}x[l] h_0[2m-l]
\end{equation}
Similarly, the samples are also passed through a high-pass filter $h_1$ and downsampled by a factor of two:
\begin{equation}
y_{high}[m] = \sum_{l=-\infty}^{+\infty}x[l] h_1[2m-l]
\end{equation}

Low-pass filter $h_0$ and high-pass filter $h_1$ can be seen as the analysis filter-bank in DWT and an example is $(5,3)$ filter-bank, where $h_0 = (-1 \quad 2 \quad 6 \quad 2 \quad -1)/8$ and $h_1=(1 \quad 2 \quad -1)/2$. The filtered samples $\mathbf{y}_{low}$  and $\mathbf{y}_{high}$ are normally named as wavelet coefficients.

\begin{figure}
\centering
\includegraphics[height=6.3cm]{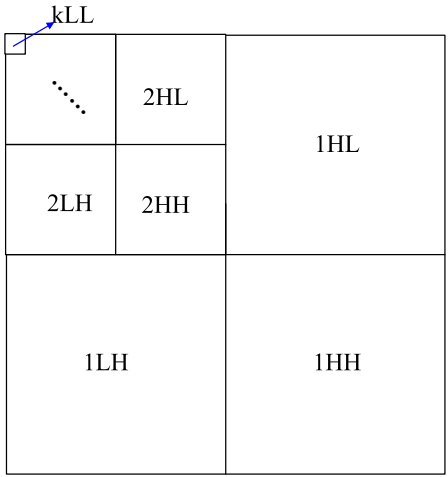}
\caption{K-level 2-D DWT decomposition.}
\label{decomposition}
\vspace{-1.5em}
\end{figure}

Then, two-dimensional DWT (2-D DWT) of image $\mathbf{I}$ can be conducted along the horizontal direction via 1-D DWT of each row of the image, and then conducted along the vertical direction via 1-D DWT of each column of the filtered and subsampled data. Such a 2-D DWT can be seen as the decomposition  resulting in four subbands of filtered and subsampled wavelet coefficients, referred to as HH (high-pass filtering in both direction), HL (high-pass filtering along horizontal direction and low-pass along vertical direction), LH (low-pass filtering along horizontal direction and high-pass along vertical direction) and LL (low-pass filtering in both direction) subbands. Then, LL subband can be further decomposed into four smaller susbands with the same decomposition manner. As seen in Figure \ref{decomposition}, k-level subbands can be achieved via decomposing the image $k$ times and finally there are $3(k-1) + 4$ subbands in total. Note that in this figure $k$LL stands for LL subbands in k-level of the 2-D DWT decomposition.

For a $k$-level 2-D DWT decomposition of image $\mathbf{I}$, the $k+1$ level progressive resolutions of the image can be reconstructed. Towards progressive resolutions, JPEG 2000 defines  resolution $0$ as the lowest resolution and  resolution $k$ as the highest resolution. Then, images at different resolution levels in JPEG 2000 can be simply achieved via the wavelet coefficients stored in different subbdands of DWT. For example, resolution $0$ of the image can be reconstructed by the wavelet coefficients of $k$LL subbands and resolution $s$ of the image can be reconstructed by the wavelet coefficients of $(k-s+1)$HL, $(k-s+1)$LH, and $(k-s+1)$HH subbands combined with the image at resolution $s-1$.

Finally, the progressive resolution recovery of JPEG 2000 can be achieved by encoding wavelet coefficients at different resolution levels into stream data $\bm{\mathbf{W}}$ with a resolution increasing order using the quantization and entropy coding. Therefore, we have  $\bm{\mathbf{W}}=[\bm{w_0}, \bm{w_1} ,\bm{w_2} ,...,\bm{w_k} ]$, where $\bm{w_i}$ is the layered stream data vector corresponding to the wavelet coefficients for reconstructing the image at level $i$, with corresponding bit rate being $w_i$ in Mbps.

\section{Progressive Fountain Codes}\label{ProFC}
\subsection{Review of Fountain Codes}\label{fc}
Fountain codes can yield an infinite number of encoding symbols on-the-fly from $K$ original symbols. In \cite{luby2002lt}, it was demonstrated that all original symbols can be recovered as long as the receiver receives any (1+$\epsilon$)$K$ encoding symbols. The overhead $\epsilon$ is possible to be very slight and when $K\longrightarrow\infty$, $\epsilon\longrightarrow0$. In a word, fountain codes have achieved capacity-approaching behavior with very low overhead.

LT codes is the first practical fountain code. At the encoder, the procedure of generating a encoding symbol is as follows. Firstly, a degree $d$ is selected randomly from a degree distribution $\bm{\Omega}(x)= \sum\nolimits_{i=1}^K{\Omega _i x^i}$, where $\Omega _i$ stands for the probability of encoding degree $i$ and satisfies $\sum\nolimits_{i=1}^K{\Omega _i}=1$. The degree distribution should be carefully designed for efficient decoding and some distributions has been proposed, such as \emph{Ideal-Soliton} distribution and \emph{Robust-Soliton} distribution\cite{luby2002lt}. Secondly, $d$ different input symbols are uniformly chosen from $K$ original symbols. Thirdly, the encoding symbol is generated by performing bitwise XOR operation on $d$ input symbols. If $d=1$, the encoding symbol is just a duplication of the unique input symbol. Lastly, the encoding symbol is transmitted. This procedure will be executed repeatedly and a potentially infinite encoding symbol stream can be generated until enough encoding symbols are collected at the client to recover all original symbols.

At the decoder, the procedure of decoding is based on belief propagation (BP)\cite{luby2002lt}. Firstly, BP process searches all receiving symbols with degree 1, which are exactly the corresponding original symbols. These symbols are stored in a buffer called \emph{ripple}. Secondly, each symbol in the ripple is recovered and at the same time other symbols out of ripple are released which may be added to the ripple if the degree becomes to 1. Since each receiving symbol is a linear combination of original symbols, of course the original symbols can be recovered by solving a linear equations called maximum likelihood decoding (ML)\cite{3GPP_MBMS}. BP decoding has lower decoding complexity and ML decoding has higher decoding efficiency.

Moreover, in order to reduce the encding/decoding overhead, Raptor codes\cite{shokrollahi2006raptor} have been proposed as an extension of LT codes with linear time encoding and decoding using a pre-coder of low-density parity-check (LDPC) codes. Our proposed progressive fountain codes also focus on Raptor codes.

\subsection{Progressive Fountain Codes}\label{ppfc}
As aforementioned, the layered stream data of JPEG 2000 images can be represented by $\bm{\mathbf{W}}=[\bm{w_0} , \bm{w_1} ,\bm{w_2} ,...,\bm{w_k} ]$. The priorities of each level decrease by $i$ from 0 to $k$. That is, data $\bm{w_0}$ with the highest priority must be decoded before all other data, while data $\bm{w_k}$ with the lowest priority have to be decoded at last. Then, in order to enhance fountain codes with UEP property, we shall propose in the following two subsections a progressive fountain codes (PFC) scheme. The PFC scheme includes two steps: rate adjustment and Raptor encoding.

\begin{figure}
\centering
\includegraphics[height=6.3cm]{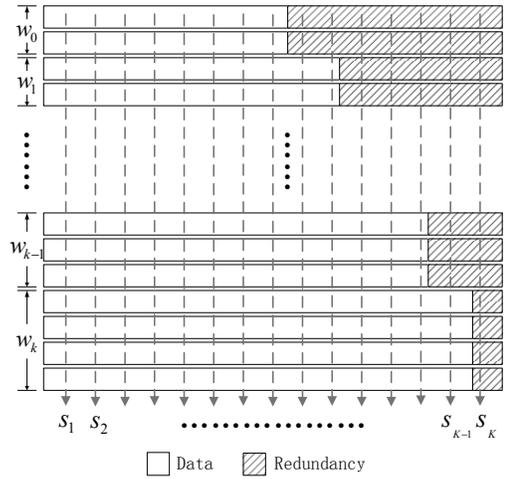}
\caption{Structure of rate adjustment and interleaving.}
\label{rateadj}
\vspace{-1.5em}
\end{figure}

\subsubsection{Rate Adjustment and Interleaving} \label{adaption}
Data at different levels should be equalized by the rate adjustment. A list of coding rate $\mathbf{r}=[r_0 , r_1 ,r_2 ,...,r_k ]$ are assigned for data at different levels according to the priority, respectively. For all $k+1$ levels it satisfies $r_0 \le  r_1  \le r_2  \le ... \le r_k$, where $r_i= \frac{{w_i}}{{n_i}}$. Note that $n_i$ is the data bit rate at the $i$th level after rate adjustment. The encoder of rate adjustment may be any maximum distance separable (MDS) codes such as Reed-Solomon (RS) code. The encoded data with various levels are then collected together as intermediate data stream $\bm{\mathbf{N}}$, whose bit rate is
\begin{equation}
n =\sum_{i=0}^{k} {\frac{{w_i }}{{r_i }}}
\end{equation}

Furthermore, the intermediate data $\bm{\mathbf{N}}$ need to be interleaved since it is oriented to the packet transmission. After that, all data at various levels are protected proportionally with their priorities in terms of rate adjustment. Rate adjustment and interleaving are illustrated in Figure \ref{rateadj}. Note that rate adjustment and interleaving are proceeded horizontally and vertically, respectively. The interleaved packets $\bm{\mathbf{S}} = [\bm{s_1} , \bm{s_2} , ... , \bm{s_K}]$ will be passed to fountain encoder as input symbols.
\subsubsection{Fountain Encoding} \label{RaptorEnc}
Standard systematic Raptor codes are performed to encode input packets $\bm{\mathbf{S}}$ into output symbols $\bm{\mathbf{O}}$ using the step of this subsection. Assume that the overhead of Raptor code is $\epsilon$, and then the overall transmission rate is $o=(1+\epsilon)n$. Here, $\epsilon$ can be dynamically adapted according to the channel condition. Then, we can obtain the total overhead $\epsilon^*$ for original input stream $\mathbf{W}$,
\begin{equation}
\label{eq4}
\epsilon^*(\mathbf{r},\epsilon) = \frac{{(1 + \epsilon)n}}{{\sum_{i=0}^k {w_i } }} - 1 = \frac{{(1 + \epsilon )\sum_{i=0}^k {\frac{{w_i }}{{r_i }} - \sum_{i=0}^k {w_i}}}}{{\sum_{i=0}^k {w_i }}}
\end{equation}

Although the Raptor encoder does not directly consider the priorities of data,  the data at higher priority level can be recovered with higher probability because of the lower adjustment coding rate compared with lower priority levels.

The advantage of our scheme is that no modification to the standard fountain codes structure is required, as to say the decoding efficiency can be guaranteed.
However, a header with rate adjustment information of each level data is necessary at the receiver, producing extra few bits which can be practically neglected for the data in large blocks.

\begin{figure}
\centering
\includegraphics[height=6.0cm]{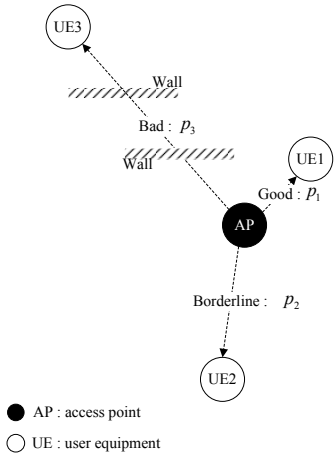}
\caption{Overview of the prototype of the proposed unequal error protected JPEG 2000 broadcast scheme.}
\label{Broadcast}
\vspace{-1.5em}
\end{figure}

\section{Prototype of the Proposed Broadcast Scheme}\label{System}
As shown in Figure \ref{Broadcast}, the prototype of the proposed unequal error protected JPEG 2000 broadcast scheme consists of a server with a wireless LAN access point (AP) and three laptop user equipments (UE). The three UEs are distributed randomly around the AP as seen in this figure, in which UE1 is the nearest while UE3 is the farthest shielded by two concrete walls. $p_i$ indicates the packet loss rate (PLR) corresponding to each UE due to the channel quality influenced by the distance and shading, and hence we have $p_1<p_2<p_3$ which will be shown in Section \ref{Result}. Layered stream data of MJPEG 2000 video are broadcasted via 802.11g at 2.4GHz by AP on a certain UDP port. Once connected to the AP, a user can join the broadcast group anytime to receive the stream data packets, being able to be decoded from anywhere in the stream since there is no inter-prediction between frame images of MJPEG 2000 video.

In contrast to the conventional broadcast schemes with equal error protection (EEP), the proposed scheme with UEP property is possible to serve more UEs with various channel conditions. In the proposed UEP broadcast scheme with PFC, the decoded resolution can be adapted to the channel quality, i.e. users with lower packet loss rate will acquire higher resolution frames while the users with higher loss rate can still be at least satisfied with lower resolution.

Assume that we have N users with PLRs satisfying $p_1 \le p_2 \le ... \le p_N$. Given total overhead $\epsilon_{0}^*$ limited by broadcast bandwidth, first of all, overhead of Raptor codes $\epsilon$ is designed to recover the lowest PLR $p_1$, ensuring complete decoding of the user with best channel quality. Then, to maximize the average decoding level of all users, we can formulate our assignment problem of the adjustment coding rates $\mathbf{r}$ as
\begin{equation}
\label{eq_max}
\begin{array}{l}
    \mathop {\max }\limits_{\mathbf{r}} AVG(\mathbf{r}) = \frac{1}{N}\sum\limits_{j = 1}^N {\sum\limits_{i = 0}^k {S(p_j ,r_i )} }  \\
    s.t. \left\{
      \begin{array}{l}
        {r_0 \le r_1 \le ... \le r_k} \\
        {\epsilon^*(\mathbf{r},\epsilon) \le \epsilon_{0}^* }
      \end{array}
    \right.
\end{array}
\end{equation}

Let $S(p_j,r_i)$ denote the recovery result at PLR $p_j$ after decoding of adjustment coding rate $r_i$ and Raptor code.
\begin{equation}
\label{eq_s}
S(p_j ,r_i) = \left\{
  \begin{array}{ll}
    1, & \hbox{\emph{if recovered with rate $r_i$;}} \\
    0, & \hbox{\emph{otherwise.}}
  \end{array}
\right.
\end{equation}

Note that $S(p_j,r_i)$ can be obtained from curves pre-generated by simulation. Thus the overall decoding level at user $j$ is $\sum\nolimits_{i = 0}^k {S(p_j ,r_i )}$. Let $P(r_i)$ denote the recovery capability of adjustment coding rate $r_i$.
\begin{equation}
\label{eq_p}
P(r_i) = \max{\{p_j | S(p_j ,r_i) = 1,1 \le j \le N\}}
\end{equation}

Since we don't have an explicit expression of $S(p_j,r_i)$, the problem can be solved by a heuristic algorithm as follows.

\begin{itemize}
\item Step 1: Adjustment coding rates $\mathbf{r}=[r_0 , r_1 ,r_2 ,...,r_k]$ are all initialized to recover at the highest PLR $p_N$. That is, for any $0 \le i \le k$, $P(r_i) = p_N$.
\item Step 2: The total overhead $\epsilon^*(\mathbf{r},\epsilon)$ is compared with $\epsilon_{0}^*$. If $\epsilon^*(\mathbf{r},\epsilon) > \epsilon_{0}^*$ then turn to Step 3, else the algorithm is terminated.
\item Step 3: For each $i$ from 1 to $k$, try to increase $r_i$ to $r_{i}^{\prime}$ individually to make $P(r_{i}^{\prime}) = p_{j+1}$, if $P(r_i) = p_j(1 \le j \le N)$. Let $\mathbf{r_{i}^{\prime}} = [r_0 , ... ,r_{i - 1}, r_{i}^{\prime}, r_{i + 1},...,r_k]$, we have the corresponding decrease of average decoding level $\Delta_i = AVG(\mathbf{r})-AVG(\mathbf{r_{i}^{\prime}})$. With $m = \arg \min{\Delta_i}$, make $\mathbf{r} = \mathbf{r_{m}^{\prime}}$ and return to Step 2.
\end{itemize}

\section{Results and Analysis}\label{Result}
In this section, experiments of MJPEG 2000 stream transmission with PFC are presented on the basis of the above prototype of broadcast scheme. We compared the results of our proposed UEP scheme and an EEP scheme with the same overall overhead $\epsilon_0^*=0.2$. In the experiments, we broadcasted a MJPEG 2000 720p high definition video stream, containing 1316 frames decomposed into five levels as shown in Table \ref{sample}. The assignment of encoding parameters of UEP scheme and the equivalent of EEP scheme are shown in Table \ref{PFC} and the block size of Raptor code is 510.

\begin{figure}
%\centering
\vspace{-1em}
\hspace{-1.5em}
\includegraphics[height=7cm]{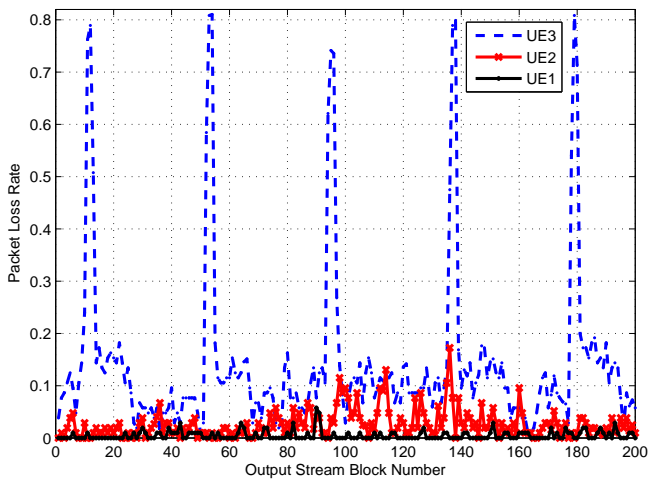}
\caption{Packet loss rates of three UEs, distributed as shown in Figure \ref{Broadcast}.}
\label{packetLoss}
\vspace{-1em}
\end{figure}

Figure \ref{packetLoss} demonstrates the packet loss rate per Raptor code block during the broadcast. The results meet our intuitive understanding of the difference of channel qualities for broadcast scheme as introduced in Section \ref{System}. Particularly, in UE3 some periodic bursts of packet loss happened, indicating that there may be a receive buffer overflow or a interference from another wireless equipment. Therefore, guaranteeing each user with various channel conditions to receive the multimedia becomes more challenging.

Table \ref{Resolution} shows the statistical results of the received resolution levels of the video for each user, in terms of $\bm{w_0}$ to $\bm{w_4}$, respectively. Note that the ``fail'' in this table means the frames being lost or decoded incorrectly. The average decoding levels of all frames are output in the last column.
It can be seen from this table that the proposed scheme outperforms EEP scheme with a much lower failing ratio all the time, in particular the channel quality is bad as for UE3 such that it will receive a continuous video stream with the proposed scheme, yet a long interruption with EEP scheme. However, the average decoding level slightly decreased when channel quality is good as for UE1. Consequently, the proposed scheme can provide all users with a progressive performance of video stream transmission, i.e. all users can be served regardless of channel conditions.

\begin{table}
\caption{The progressive resolutions and bit rates of MJPEG 2000 encoding for the sample video stream.}
\label{sample}
\centering
\scriptsize
\begin{tabular}{ c c c c c c }%>{\centering}m{1cm} >{\centering}m{0.9cm} >{\centering}m{1cm} >{\centering}m{1.2cm} >{\centering}m{1.2cm}
  \hline
  \vspace{0.5em}
                &$\bm{w_0}$&$\bm{w_1}$&$\bm{w_2}$&$\bm{w_3}$&$\bm{w_4}$\\
  \vspace{0.5em}
  Resolution &$80\times45$\hspace{-1em}&\hspace{-1em}$160\times90$\hspace{-1em}&\hspace{-1em}$320\times180$\hspace{-1em}&\hspace{-1em} $640\times360$\hspace{-1em}&\hspace{-1em}$1280\times720$\\
  \vspace{0.5em}
  Bit rate(Mbps)& 1.603 & 3.200 & 8.326 & 9.773 & 3.544 \\
  \hline
\end{tabular}
\vspace{-1em}
\end{table}

\begin{table}
\caption{The assignment of encoding parameters of UEP scheme and the equivalent of EEP scheme.}
\label{PFC}
\centering
\begin{tabular}{ c c c c c c c }
  \hline
  \vspace{0.5em}
      & $r_0$ & $r_0$ & $r_2$ & $r_3$ & $r_4$ & $\epsilon$\\
  \vspace{0.5em}
  UEP & $\frac{127}{255}$ & $\frac{191}{255}$ & $\frac{223}{255}$& $\frac{239}{255}$ & $\frac{247}{255}$ & $\frac{12}{510}$\\
  \vspace{0.5em}
  EEP & $\frac{217}{255}$ & $\frac{217}{255}$ & $\frac{217}{255}$& $\frac{217}{255}$ & $\frac{217}{255}$ & $\frac{12}{510}$\\
  \hline
\end{tabular}
\vspace{-1em}
\end{table}

\begin{table}
%\vspace{-.5em}
\caption{Statistical experimental results of resolution levels of the received video for the UEs of Figure \ref{Broadcast}.}
\scriptsize
\label{Resolution}
\centering
\vspace{-0.5em}
\begin{tabular}{ c c|c c c c c c|c }
  \hline
  \multicolumn{2}{c|}{}& \multicolumn{6}{c|}{Proportions of Resolution Levels of Received Video($\%$)}& Average \\\cline{3-8}
  \multicolumn{2}{c|}{}& fail &   $\bm{w_0}$  &   $\bm{w_1}$  & $\bm{w_2}$    &  $\bm{w_3}$   & $\bm{w_4}$  & Level\\
  \hline
  \multirow{2}*{UE1}\hspace{-1em} & UEP &      0      &\bf{0.38} &\bf{3.42} &\bf{10.79} &\bf{2.51} &   82.90   &   4.64     \\
  %\hline
      & EEP & \bf{1.44} &   0.08   &   0.30   &   0.76   &   0.08   &\bf{97.34} &\bf{4.90}   \\
  \hline
  \multirow{2}*{UE2}\hspace{-1em} & UEP &    0.15   &   0.61   &\bf{2.43} &   22.11   &\bf{5.62} &   69.07   &\bf{4.40}   \\
  %\hline
      & EEP & \bf{13.60} &\bf{0.76} &   1.90   &   2.66   &   0.68   &\bf{80.40} &   4.17     \\
  \hline
  \multirow{2}*{UE3}\hspace{-1em} & UEP &    16.72   &\bf{50.38} &\bf{23.86} &\bf{8.13} &   0.23   &   0.68   &\bf{1.27}   \\
  %\hline
      & EEP & \bf{81.23} &   1.44   &   3.80   &   3.88   &\bf{0.84} &\bf{8.81} &   0.68     \\
  \hline
\end{tabular}
\vspace{-1em}
\end{table}

\section{Conclusions}\label{conclusion}
This paper has presented a JPEG 2000 broadcast scheme with progressive fountain codes, based on unequal error protection for layered video streams. With the proposed progressive fountain codes in such a scheme, it was demonstrated that our broadcast scheme is capable of serving various users adaptively with different resolution levels according to the channel quality. When the transmitted packets are at an extremely high loss rate, video streams can still be decoded with lower resolution by the proposed scheme while the conventional EEP schemes fail to decode. Experimental results suggest the superiority of our scheme versus EEP schemes, indicating that our scheme is insensitive to the channel quality and thus will be an efficient solution to multimedia broadcast systems in the future.

\vspace{-0.5em}
\section*{Acknowledgment}
This research is supported by the National Basic Research Program of China (2007CB310601).

\bibliographystyle{IEEEtran}
\bibliography{IEEEfull,CZBib}

\end{document}